\documentstyle[12pt,aaspp]{article}

\def\SZ{{\rm Sunyaev-Zel'dovich}}
\def\cl{{\rm CL~0016+16}}
\def\rxold{{\rm RX~J0018.3+1618}}
\def\rxnew{{\rm RX~J0018.8+1602}}
\def\etal{{\rm et al.}}
\def\asca{{\it ASCA}}
\def\rosat{{\it ROSAT}}
\def\ginga{{\it Ginga}}

\def\myarcmin{^\prime\mskip-5mu}
\def\myarcsec{^{\prime\prime}\mskip-5mu}
\def\lsim{\hbox{\raise.35ex\rlap{$<$}\lower.6ex\hbox{$\sim$}\ }}
\def\gsim{\hbox{\raise.35ex\rlap{$>$}\lower.6ex\hbox{$\sim$}\ }}

\begin{document}

\received{24 June 1997}
\accepted{17 November 1997}

\slugcomment{To appear in {\it The Astrophysical Journal}}

\title{Another X-ray--Discovered Poor Cluster of Galaxies Associated
with CL~0016+16\footnote{Optical observations reported here were
  obtained at the Multiple Mirror Telescope, a joint facility of the
  Smithsonian Institution and the University of Arizona.}}

\author{John P.~Hughes}
\affil{jph@physics.rutgers.edu\\ 
Department of Physics and Astronomy, Rutgers University,\\ 
136 Frelinghuysen Road, Piscataway, NJ 08854-8019}
\centerline{and}
\author{Mark Birkinshaw}
\affil{mark.birkinshaw@bristol.ac.uk\\ Department of Physics, University
of Bristol, Bristol, BS8 1TL, UK}

\keywords{
 cosmology: observations ---
 galaxies: clusters: individual (\cl, \rxold, \rxnew) ---  
large-scale structure of universe
}

\begin{abstract}

We report optical spectroscopic observations of \rxnew, a \rosat\
X-ray source proposed to be an intermediate redshift cluster of
galaxies.  Our observations confirm the identification of \rxnew\ and
provide measurements of its mean radial velocity ($z=0.5406\pm0.0006$)
and velocity dispersion ($\sigma_p = 200^{+110}_{-80}$ km s$^{-1}$).
This is the second poor cluster that has been found to be a companion
to \cl\ ($z=0.5455$), the other one being \rxold\ ($z=0.5506$).  The
0.2--2 keV band source-frame X-ray luminosity summed over both
companion clusters is $5\times 10^{44}$ ergs s$^{-1}$, which is a
significant fraction $\sim$23\% of the X-ray luminosity of the main
cluster.  The companions are located at angular distances of
10$^\prime$ to 25$^\prime$ (minimum physical scales of 5 Mpc to 12
Mpc) from \cl\ and we propose that they represent a new large-scale
component of the X-ray emission from clusters of galaxies.  Similar
low X-ray luminosity poor clusters surrounding nearby Abell clusters
can explain the excess power observed in the angular cross-correlation
function between Abell clusters and the X-ray background on inferred
physical scales of 14--20 Mpc (Soltan et al.~1996).

\end{abstract}

\section{Introduction}

In this article, we continue our studies of the environment of the
moderately distant ($z=0.5455$) rich cluster of galaxies \cl.  This
cluster was the prime target of a deep (43 ks) \rosat\ X-ray pointing
that has been used in interpreting the cluster's \SZ\ effect (Hughes
\& Birkinshaw 1997) and estimating its gravitational mass (Neumann \&
B\"ohringer 1997). We have been following-up selected serendipitous
sources in the imaging data (there are $\sim$80 field sources detected
by the \rosat\ standard processing) in order to investigate the large
scale structure or supercluster believed to exist near \cl\ (Koo 1981,
1986; Connolly et al.~1996).  The length of the observation, plus the
high sensitivity of the \rosat\ position sensitive proportional
counter (PSPC), offsets to some extent the low source fluxes of
objects at these redshift.  On the other hand the large redshift
provides a clear advantage for studying structure: the roughly
1$^\circ$ field of view of the PSPC corresponds to large physical
scales at the cluster's redshift: 30 Mpc or so.  Throughout this
article we assume $H_0 = 50$ km s$^{-1}$ Mpc$^{-1}$, $q_0 = 0.15$.

Results to date corroborate the approach of using X-ray--selection to
probe large scale structure.  Previously we reported the discovery,
based on the \rosat\ data, of a poor cluster, \rxold, some
$9\myarcmin.5$ (4.6 Mpc) southwest of \cl, which our optical
spectroscopy revealed to be at nearly the same redshift (Hughes,
Birkinshaw, \& Huchra 1995).  Below we present new optical
observations of a second X-ray source in this field, \rxnew, proposed
to be yet another galaxy cluster at roughly the same redshift
(Connolly et al.~1996).  Our observations confirm its identification
and provide the first accurate measurements of its redshift and
velocity dispersion.  It is indeed another companion to \cl, located
$\sim$25$^\prime$ (12 Mpc) nearly due south.

We consider the existence of these two companions to \cl\ in light of
the recently discovered extended, low surface brightness X-ray
``halos'' apparently associated with Abell clusters of galaxies
(Soltan et al.~1996).  The evidence for these X-ray halos comes from a
cross-correlation between the positions of Abell clusters and the
X-ray background as measured by the \rosat\ All-Sky Survey (RASS)
that was found to show significant amplitude on angular scales of
1$^\circ$ to 3$^\circ$. Abell clusters are known to be
highly clustered (e.g., Nichol, Briel, \& Henry 1994) and are strong
X-ray emitters, nevertheless Soltan et al.\ claim that the signal they
detect is a factor of three larger than the correlation function
predicted from the known clustering properties of Abell clusters
themselves, signifying that an additional, new component of sources,
spatially correlated with Abell clusters and displaying soft X-ray
emission, is required.  The X-ray luminosity of the extended component
is $\sim$$10^{44}$ ergs s$^{-1}$ in the \rosat\ band.  Soltan et al.\
offered two explanations for the halos: X-ray emission from either (1)
a number of discrete sources associated with groups or poor clusters,
or (2) a truly diffuse component of hot gas extended over tens of
Megaparsecs. Note that the latter component has an inferred X-ray
surface brightness that is about 2 orders of magnitude less than the
soft X-ray background and thus it is unlikely that individual examples
could be imaged directly by \rosat.

\par

As we show below, the locations of the new clusters relative to \cl\,
and their X-ray luminosities and temperatures are broadly consistent
with what is needed to explain the correlation function results in
terms of discrete sources.  This conclusion is further strengthened
by the significant number of low redshift groups of clusters that may
represent nearby examples of this phenomenon.

\section{Observations}
\subsection{X-ray}

The reprocessed data from a 43 ks \rosat\ PSPC observation were used
to study the X-ray properties of the proposed new cluster.  \rxnew\ is
approximately 25$^\prime$ off-axis, where the imaging resolution of
the PSPC has a broad core ($\sim$2$^\prime$ FWHM) due to the off-axis
aberrations of the \rosat\ X-ray mirrors, which makes detailed
analysis of the spatial structure difficult. Another complication
arises from the presence of a second X-ray source about 3$^\prime$ to
the southwest.  Both sources can be seen clearly in Figure 1 (Plate
L00) where the contour map shows the X-ray data. Note that the
absolute accuracy of PSPC source positions was verified to be better
than 2$^{\prime\prime}$ based on agreement between the X-ray and
optical positions of five point sources (Hughes et al.~1995).

We fit the X-ray imaging data within a 4$^\prime$ radius of \rxnew\
using a model consisting of background, a point source, and an
isothermal-$\beta$ model cluster profile of the form $\Sigma =
\Sigma_0 [1 + (\theta/\theta_C)^2]^{-3\beta+0.5}$ (Cavaliere \&
Fusco-Femiano 1976).  For background we used the background image
provided as part of the standard processing. The data strongly
require, at about the 3-$\sigma$ level, that the northern source be
extended, although the core radius parameter in the cluster model,
$\theta_C$, is not well-constrained due to the large image blur at the
source's off-axis position. With $\theta_C$ fixed to the value
$0\myarcmin.44$ found earlier for \rxold\ (Hughes et al.~1995), we 
find $\Sigma_0 \approx 1.1\times 10^{-2}$ counts s$^{-1}$
arcmin$^{-2}$ and $\beta \approx 0.75$ for the candidate new cluster.
The position of the cluster model (from maximum likelihood fits to the
image) is 00$^{\rm h}$18$^{\rm m}$47.9$^{\rm s}$,
16$^\circ$02$^\prime$22$^{\prime\prime}$ (J2000) with an uncertainty
of about 5$^{\prime\prime}$ in each coordinate.  In the spectral
analysis presented next, we were careful to select the spectral
extraction regions so that the X-ray emission from the two sources was
kept separate.

The point-like X-ray source to the south (at 00$^{\rm h}$18$^{\rm
m}$42.6$^{\rm s}$, 15$^\circ$59$^\prime$37$^{\prime\prime}$) has a
count rate within a radius of $1\myarcmin.5$ of $(5.8\pm 0.6)\times
10^{-3}$ PSPC counts s$^{-1}$.  Although there are too few counts
($\lsim$200) to do a detailed spectral analysis, the hardness ratios
are consistent with a low column density ($N_{\rm H} \lsim 10^{19}$
cm$^{-2}$) and a spectrum that cuts-off above $\sim$1 keV, similar to
that expected from a Galactic star. The unabsorbed X-ray flux is
$\sim$$3.3\times 10^{-14}$ ergs cm$^{-2}$ s$^{-1}$.

The northern source, i.e., the cluster candidate, has an observed
count rate (within a radius of $1\myarcmin.8$) of $(9.4\pm0.8)\times
10^{-3}$ PSPC counts s$^{-1}$ based on a total of approximately 310
counts.  The spectrum can be fit well by a thermal plasma model using
the Galactic column density ($N_{\rm H} = 4.1\times 10^{20}$
cm$^{-2}$) and fractional elemental abundances of \onequarter\ solar.
Anticipating the results in \S 2.3 below, we set the redshift
to 0.5406 and obtained a best-fit source-frame temperature of
$kT=1.4^{+0.7}_{-0.3}$ keV (1 $\sigma$ uncertainty), in good agreement with
Connolly et al.~(1996).  The unabsorbed X-ray flux is $1.5\times
10^{-13}$ ergs cm$^{-2}$ s$^{-1}$ over the 0.2--2 keV band.

\subsection{Optical Imaging}

The candidate cluster was observed at the Whipple Observatory 1.2-m
telescope on 1996 July 17 under fair to poor conditions for 1800 s in
$R$.  The focal plane detector CCD was binned to produce an effective
pixel scale of $0\myarcsec.64$.  Using standard {\sc iraf} tasks, we
bias-subtracted the data, eliminated cosmic rays and bad pixels, and
flat-fielded using dome flats.  The positions of four {\it HST} guide
stars were used to register the image to the celestial sphere; a
positional uncertainty of $\lsim$$0\myarcsec.5$ was obtained.  Stellar
images were about $1\myarcsec.8$ [FWHM], which is not unusual for this
telescope.

\par

Figure 1 (Plate L00) shows a portion of the $R$-band image with
contours of X-ray emission superposed. Numerous faint galaxies are
associated with the diffuse X-ray emission, with a dominant bright
galaxy (labeled 4) within 7$^{\prime\prime}$ of the center of the
X-ray emission. {\sc daophot} was run on the frame to detect objects
and of order 300 were found. The sharpness parameter was used to
discriminate between stars and galaxies for the follow-up optical
spectroscopy.

\par

Toward the south there is a bright ($m_V = 8.6$) star (HD~1450) of
spectral type F5 that is positionally coincident with the southern
X-ray source.  Using the flux quoted above, we estimate an X-ray to
optical flux ratio for the star of $\log f_X/f_V \sim -4.7$ (using the
relationship in Stocke et al.~1991) that is in the range expected
for Galactic stars of spectral type B-F.  We conclude that HD~1450 is
the counterpart to the southern X-ray source.

\subsection{Optical Spectroscopy}

We obtained optical spectra at the Multiple Mirror Telescope on 1996
November 12 and 13 using the Red Channel with aperture plates
(Fabricant, McClintock, \& Bautz 1991).  Four aperture plates were
machined several weeks before the run with slitlets cut for seven
candidate galaxies on each plate. Our spectral resolution was $\sim$15
\AA\ (FWHM) and all spectra included the wavelength range 5500 -- 8500
\AA. Multiple exposures (2700 s duration) of each plate were taken for
total integration times ranging from 5400 s to 13500 s.  He-Ne-Ar lamp
exposures were done before and after each exposure and dome flats were
done at the beginning or end of the night.

\par

Standard {\sc iraf} routines were used for cosmic ray removal,
bias-level subtraction, flat-fielding, and to trace and extract the 
one-dimensional spectra.  Wavelength calibration used $\sim$45 lines
from the comparison lamp spectra fit to a cubic spline
(root-mean-square residuals were $\sim$1.0 \AA).  We verified the
wavelength of the 6300.2 \AA\ night sky line in our spectra when the
data were extracted without sky subtraction.  Differences were $\sim$1
\AA\ and our derived redshifts were corrected for this shift ($\Delta
z \sim 0.0002$). To derive recessional velocities our reduced
spectra were cross-correlated (task {\sc fxcor}) over the observed
wavelength range 5600--7500 \AA\ with a spectrum of the galaxy NGC
4486B.

\par

The spectrum of the bright central galaxy (\# 4) is shown in Figure 2.
The 4000 \AA\ break, as well as absorption features due to Ca {\sc ii}
H and K, the G-band, and Mg $b2$, are all clearly visible.  Our
redshift $z=0.5413\pm0.0007$ is consistent with the recent value of
$0.541\pm0.004$ presented by Connolly et al.~(1997). Strong
correlations ($R>4.0$) were obtained for the 15 galaxies listed in
Table 1, which turn out to be all absorption line objects. We point
out object g in the table, which is the radio galaxy 54W084
(Windhorst, Koo, \& Kron 1984). Our redshift is consistent with a less
accurate value by R.~Windhorst from an unpublished low
signal-to-noise, low resolution Palomar spectrum.

\par

We were unable to identify or determine redshifts for the remaining 13
objects on the various plates.  Our ``success'' rate was highly
correlated with the weather conditions. On the first night (clear to
hazy conditions) 10 out of 14 of our observed spectra resulted in
redshift measurements, while on the second night (hazy to partly
cloudy conditions) only 5 out of 14 did.  Of the 21 objects observed
within $1\myarcmin.6$ of the cluster center, 12 were identified as
galaxies and 8 are members of the cluster. On the other hand, all of
the identified galaxies beyond $1\myarcmin.4$ have radial velocities
that are inconsistent with cluster membership.

\par

The formal velocity uncertainties generated by {\sc fxcor} from the
width of the correlation function vary from galaxy to galaxy over the
range 100 km s$^{-1}$ to 250 km s$^{-1}$.  In addition we include an
estimate of the velocity uncertainty due to sky subtraction, obtained
by carrying out the reduction twice using slightly different
parameters of the fitting function for the sky and taking the
difference between the resulting redshift values as the additional
uncertainty. In general this quantity is comparable in magnitude to
the previous uncertainty; we find values ranging from 80 km s$^{-1}$
up to 330 km s$^{-1}$.  We also include an uncertainty of 60 km
s$^{-1}$ due to wavelength calibration and an uncertainty of 90 km
s$^{-1}$, due to variation of the parameters of the cross-correlation,
e.g., the order of the continuum fits and the wavelength range over
which the cross-correlation was done. The final uncertainty quoted in
Table 1 is the root-sum-square of the individual components.

\section{Discussion}

The mean redshift of the new cluster based on the eight galaxies
indicated as members in Table 1 is $0.5406\pm 0.0006$ and the radial
component of its velocity dispersion is $200^{+110}_{-80}$ km
s$^{-1}$ (1 $\sigma$ uncertainties), calculated using the formulae in
Danese, De Zotti, \& di Tullio (1980). Although galaxies b, f, and g
in the lower part of the table have redshifts near 0.5, they are
unlikely to be members of the cluster because they lie at the
outskirts of the cluster (more than $1\myarcmin.5$ from the central
galaxy), and their redshifts are more than 8 $\sigma$ away from the
mean of the cluster.  They are possibly related to the sheet of
galaxies identified by Connolly et al.~(1996).

The X-ray luminosity of \rxnew\ is $3 \times 10^{44}$ ergs s$^{-1}$
over the 0.2--2 keV band in the source frame. From the measured
temperature, the imaging results, and the standard assumption of a
symmetric, isothermal cluster in hydrostatic equilibrium, we estimate
the cluster's mass within a 2 Mpc radius to be $3 \times 10^{14}$
$M_\odot$.  The ratio of gas mass to total mass within this same
radius is 33\%.  These values are similar to those of the other poor
cluster, \rxold\ (Hughes et al.~1995) associated with \cl\ and in
Table 2 we present a comparison between the observed and physical
characteristics of the two new clusters. (Note that the temperature of
\rxold\ was determined for this article from a joint fit of \rosat\
PSPC and \asca\ GIS data.)  The luminosity of \cl\ is $2\times
10^{45}$ ergs s$^{-1}$, which is 4 times the combined X-ray emission
from the companions. Finally we note for completeness that the X-ray
luminosities of the companion clusters are quite a bit larger than the
average X-ray luminosity of nearby galaxy groups ($L_X \sim 10^{41}-
10^{42}$ ergs s$^{-1}$, Mulchaey et al.~1996).

Energy considerations were used by Hughes et al.~(1995) to show that 
\cl\ and \rxold\  were not unlikely to form a bound
system: on the assumption of random projections of the separation
vector and relative velocity vector of these clusters, and using the
mass estimates out to radii of 2~Mpc, the {\it a priori} probability
that these two clusters form a bound system is about 38\%. We have
repeated this calculation taking into account the extra kinetic and
potential energies contributed by \rxnew. The {\it a priori}
probability that all three clusters form a bound system can be shown
to be about 11\%. This is smaller than previously calculated, because
\rxnew\ contributes less to the potential than the kinetic
energy --- it is rapidly moving relative to, and distant from,
the center of mass of the composite system for most choices of
projection angles.  If \cl\ is taken, instead, to have a radius of
5~Mpc, then the mass increases by a factor 2.5 and the probability
that the entire system is bound rises to about 30\%. 
These {\it a priori} probability estimates are likely to be
pessimistic about the chances that the overall system is bound, since
they ignore the indications of further clumped masses in the field of
\cl\ (based on optical material: Connolly et al.~1996).

If we assume that the projection angles of the companion clusters
relative to \cl\ are less than $60^\circ$ (i.e., the three clusters
lie within a sphere of radius $\sim$25 Mpc centered on \cl), then a
conservative lower limit to the local space density of clusters is $N
\sim 5\times 10^{-5}$ Mpc$^{-3}$.  This is 2 orders of magnitude
larger than the space density of Abell clusters with $L_X \gsim
2\times 10^{44}$ ergs s$^{-1}$ from a complete sample of clusters over
several hundred square degrees at high Galactic latitude (Briel \&
Henry 1993).

Notwithstanding this large local space density of clusters, it is our
opinion that the \cl\ system is not particularly exceptional but may
merely represent an intermediate redshift example of essentially
similar nearby systems.  We offer two pieces of evidence in support of
this. One is the high frequency of pairs observed in the X-ray
flux-limited sample of low redshift ($z<0.2$) clusters from Lahav et
al.\ (1989).  The other is the presence of extended, low surface
brightness X-ray haloes associated with Abell clusters mentoned in the
Introduction.  We discuss each of these below.

Out of the 55 clusters in the all-sky sample of the brightest X-ray
clusters (Lahav et al.\ 1989, Edge et al.\ 1990), there are six close
pairs with physical separations of roughly 20 Mpc or less.  Since both
clusters must appear in the list, this number is clearly a lower
limit to the fraction of such pairs. A better estimate can be obtained
by searching for companions from the \rosat\ X-ray brightest Abell
clusters (the XBACs sample, Ebeling et al.\ 1997) which goes to a much
lower flux limit (nearly a factor of 10). Of the 46 clusters in the
Edge et al.\ sample at high Galactic latitude ($b > 20^\circ$, which
is also the XBACs limit), we find that as many as 19 have X-ray
emitting companions nearby (i.e., within physical separations of 25
Mpc or  less), for a fraction of clusters with close companions of 41\%.
Again this is a lower limit since the XBACs sample was derived from
the (optically-selected) Abell clusters.

In the XBACs sample itself there are 30 systems that contain two,
three, four, or even five individual clusters in close proximity (25
Mpc).  The median $L_X$ of the sample consisting of the dominant
cluster in each system is $3.5\times 10^{44}$ erg s$^{-1}$, while the
median luminosity of all the companion clusters (of which there are
41) is $1.9\times 10^{44}$ erg s$^{-1}$, which is nearly identical to
the luminosity of \rxold\ and \rxnew. If, for each individual system,
one compares the companion's X-ray luminosity (summed over all
companions in systems with more than one) to that of the dominant
cluster, one finds that the ratio varies over a large range, 0.17 to
1.5, as plotted in Fig.~3.  The filled circle shows the \cl\ system.
These results firmly establish that the \cl\ system is entirely
consistent with the close cluster groupings from the low redshift
XBACs sample.

The second piece of evidence for structure on ten Megaparsec scales
associated with nearby clusters of galaxies comes from a
cross-correlation between the angular positions of Abell clusters and
the intensity of the soft X-ray background as measured by \rosat\
(Soltan et al.~1996).  To explain their cross-correlation results
Soltan et al.\ need a new component of X-ray emission from structure
on scales 20--30 times larger than the size of the cluster itself, or
physical scales of $\sim$20 Mpc.  The spectral character of the emission
needs to be softer than the X-ray background in general, due to limits
set by the \ginga\ satellite over the higher energy 4--12 keV band
(from the auto-correlation function of the X-ray background, Carrera
et al.~1991, 1993).  Sources with little emission above $\sim$4 keV
are required and thermal spectra with $kT \lsim 5$ keV would be
consistent with this limit. And as mentioned in the Introduction the
mean luminosity of the extended component is $L_X\sim 10^{44}$ ergs
s$^{-1}$. 

One possible explanation for this new component of X-ray emission is
the hot gas that may be contained in filaments connecting galaxy
clusters.  Such filaments are a common feature of numerical models of
structure formation in the Universe; there is also observational
support for their existence from galaxy redshift surveys (e.g., Da
Costa \etal\ 1994). Recently, however, Briel \& Henry (1995) derived a
limit on the X-ray emission from such filaments that is significantly
less than what would be needed to explain the excess correlation
power.  On the other hand, in all direct observational aspects:
spatial scale, low X-ray temperature, and soft X-ray luminosity, the
companion clusters to \cl\ satisfy the observational requirements
demanded of the extended halo emission component. Our results
therefore strongly suggest that the origin of the extended halos
around Abell clusters should be interpreted as arising from physically
associated discrete sources, specifically poor clusters or groups of
galaxies, rather than a truly diffuse extended halo of hot
gas. Whether on the average the companion clusters are few and have
luminosities consistent with poor clusters ($\sim$10$^{44}$ ergs
s$^{-1}$) or are more common but less luminous, like groups, remains
to be seen.

\vspace{0.5in}

We thank Ale Milone for taking the R-band images of \rxnew\ and for
her efforts as MMT operator during the run in November. Thanks are
extended to Paul Callanan and Mike Garcia for sharing their 1.2-m time
and we acknowledge the generosity of the CfA TAC for awarding us MMT
time. We thank the referee, Rogier Windhorst, for his helpful
comments.  This research was partially supported by a PPARC grant to
MB and NASA Long Term Space Astrophysics Grant NAG5-3432 to JPH.

\clearpage

\figcaption[]{R-band image from the Mount Hopkins 1.2 m telescope of a
portion of the field containing the X-ray--discovered cluster, \rxnew,
overlaid with contours of constant X-ray surface brightness.  Contour
values increase in multiplicative steps of 1.476 from a minimum value
of $2.0\times 10^{-4}$ PSPC counts s$^{-1}$ arcmin$^{-2}$ to a maximum
value of $1.4\times 10^{-3}$ PSPC counts s$^{-1}$ arcmin$^{-2}$. The 8
galaxies with measured redshifts in the range $0.538-0.542$ (Table 1)
are numbered. Galaxies labelled a to g have spectroscopic redshifts
outside this range and are believed to foreground or background
objects. Coordinates are quoted in epoch J2000. \label{fig1}}

\figcaption[]{Optical spectrum of the brightest galaxy in the new
cluster \rxnew.  The 4000\AA\ break, as well as absorption lines
characteristic of late type stellar spectra are clearly visible.  We
indicate with filled circles features in the galaxy spectrum due to Ca
{\sc ii} K and H, the G-band, Fe {\sc i} $\lambda$4383.6, Mg {\sc i}
$\lambda$5172.7 (Mg $b2$), and Fe {\sc i} $\lambda$5269.6.  Tellaric
absorption from the O$_2$ B and A bands are noted with $\oplus$
symbols.  For this display only, the spectrum was smoothed with a
15\AA\ [FWHM] gaussian, which is approximately the spectral
resolution. \label{fig2}}

\figcaption[]{The X-ray luminosity in the 0.1--2.4 keV \rosat\ band of
the dominant cluster (horizontal axis) compared to the summed
luminosity of companion clusters within a radius of 25 Mpc.  The
vertical axis plots the luminosity ratio of the companions to the
dominant cluster. Clusters from the low redshift XBACs sample are
shown as crosses, and the filled circle shows the intermediate
redshift \cl\ system. \label{fig3}}

\clearpage
\begin{deluxetable}{cccc}
\tablecaption{Spectroscopic Results}
\tablewidth{0pt}
\tablehead{
 & \multicolumn{2}{c}{Position (J2000)} \nl\cline{2-3}
\colhead{n} & \colhead{$\alpha$} & \colhead{$\delta$} & \colhead{$z$}
} 
\startdata
\multicolumn{4}{c}{\rxnew\ Cluster Members}\nl
1 & $0^{\rm h}18^{\rm m}52.9^{\rm s}$ &
     $16^\circ02^\prime 08^{\prime\prime}$   & 0.5392 $\pm$ 0.0007 \nl
2 & $\phantom{0^{\rm h}18^{\rm m}}50.7^{\rm s}$ &
     $\phantom{16^\circ}01^\prime42^{\prime\prime}$ 
     & 0.5419 $\pm$ 0.0007 \cr
3 & $\phantom{0^{\rm h}18^{\rm m}}48.2^{\rm s}$ &
     $\phantom{16^\circ}03^\prime10^{\prime\prime}$ 
     & 0.5384 $\pm$ 0.0007 \cr
4 & $\phantom{0^{\rm h}18^{\rm m}}47.7^{\rm s}$ &
     $\phantom{16^\circ}02^\prime15^{\prime\prime}$ 
     & 0.5413 $\pm$ 0.0007 \cr
5 & $\phantom{0^{\rm h}18^{\rm m}}47.3^{\rm s}$ &
     $\phantom{16^\circ}02^\prime19^{\prime\prime}$ 
     & 0.5416 $\pm$ 0.0010 \cr
6 & $\phantom{0^{\rm h}18^{\rm m}}45.8^{\rm s}$ &
     $\phantom{16^\circ}02^\prime09^{\prime\prime}$ 
     & 0.5396 $\pm$ 0.0012 \cr
7 & $\phantom{0^{\rm h}18^{\rm m}}43.9^{\rm s}$ &
     $\phantom{16^\circ}02^\prime16^{\prime\prime}$ 
     & 0.5418 $\pm$ 0.0008 \cr
8 & $\phantom{0^{\rm h}18^{\rm m}}42.0^{\rm s}$ &
     $\phantom{16^\circ}01^\prime59^{\prime\prime}$ 
     & 0.5406 $\pm$ 0.0009 \cr
\nl
\multicolumn{4}{c}{Probable Non-members}\nl
a & $0^{\rm h}18^{\rm m}57.7^{\rm s}$ &
     $16^\circ01^\prime 37^{\prime\prime}$   & 0.1594 $\pm$ 0.0006 \nl
b & $\phantom{0^{\rm h}18^{\rm m}}55.5^{\rm s}$ &
     $\phantom{16^\circ}02^\prime12^{\prime\prime}$ 
     & 0.5509 $\pm$ 0.0007 \cr
c & $\phantom{0^{\rm h}18^{\rm m}}46.7^{\rm s}$ &
     $\phantom{16^\circ}02^\prime41^{\prime\prime}$ 
     & 0.3864 $\pm$ 0.0007 \cr
d & $\phantom{0^{\rm h}18^{\rm m}}45.8^{\rm s}$ &
     $\phantom{16^\circ}01^\prime29^{\prime\prime}$ 
     & 0.6243 $\pm$ 0.0009 \cr
e & $\phantom{0^{\rm h}18^{\rm m}}45.2^{\rm s}$ &
     $\phantom{16^\circ}00^\prime49^{\prime\prime}$ 
     & 0.2097 $\pm$ 0.0006 \cr
f & $\phantom{0^{\rm h}18^{\rm m}}43.8^{\rm s}$ &
     $\phantom{16^\circ}03^\prime46^{\prime\prime}$ 
     & 0.5484 $\pm$ 0.0010 \cr
g\tablenotemark{a} & $\phantom{0^{\rm h}18^{\rm m}}41.4^{\rm s}$ &
     $\phantom{16^\circ}02^\prime10^{\prime\prime}$ 
     & 0.5491 $\pm$ 0.0012 \cr
\tablenotetext{a}{Radio galaxy 54W084 (Windhorst, Koo, \& Kron 1984)}
\enddata
\end{deluxetable}

\clearpage
\begin{deluxetable}{lcc}
\tablecaption{Clusters Associated with CL~0016+16\tablenotemark{\,\, a}\,}
\tablewidth{0pt}
\tablehead{
Parameter & \rxold\ & \rxnew 
}
\startdata
$z$                      & $0.5506\pm0.0012$   & $0.5406\pm0.0006$ \cr
$\sigma_P$ (km s$^{-1}$)  & $540^{+260}_{-120}$ &  $200^{+110}_{-80}$ \cr
$kT$  (keV)              & $3.1^{+2.8}_{-1.4}$ & $1.4^{+0.7}_{-0.3}$ \cr
$F_X$\tablenotemark{b}\, (erg cm$^{-2}$ s$^{-1}$) & $1.1\times 10^{-13}$
& $1.5\times 10^{-13}$ \cr
$L_X$\tablenotemark{c}\, (erg s$^{-1}$)     & $2\times 10^{44}$ &
$3\times 10^{44}$ \cr 
$M_{\rm gas}(<2\,{\rm Mpc})$ ($M_\odot$) & $9\times 10^{13}$ 
  & $1\times 10^{14}$ \cr
$M_{\rm tot}(<2\,{\rm Mpc})$ ($M_\odot$) & $5\times 10^{14}$
 & $2\times 10^{14}$ \cr
Distance from \cl    \cr
\quad Angular                  & $9\myarcmin.5$ & $24\myarcmin.2$ \cr
\quad Projected linear (Mpc) & 4.6 & 11.7 \cr
\tablenotetext{a}{Uncertainties quoted at the 1 $\sigma$ confidence level}
\tablenotetext{b}{Unabsorbed flux over 0.2--2 keV band (observed frame)}
\tablenotetext{c}{Over 0.2--2 keV band (source frame)}
\enddata
\end{deluxetable}

\singlespace

\clearpage

\begin{figure}
\plotfiddle{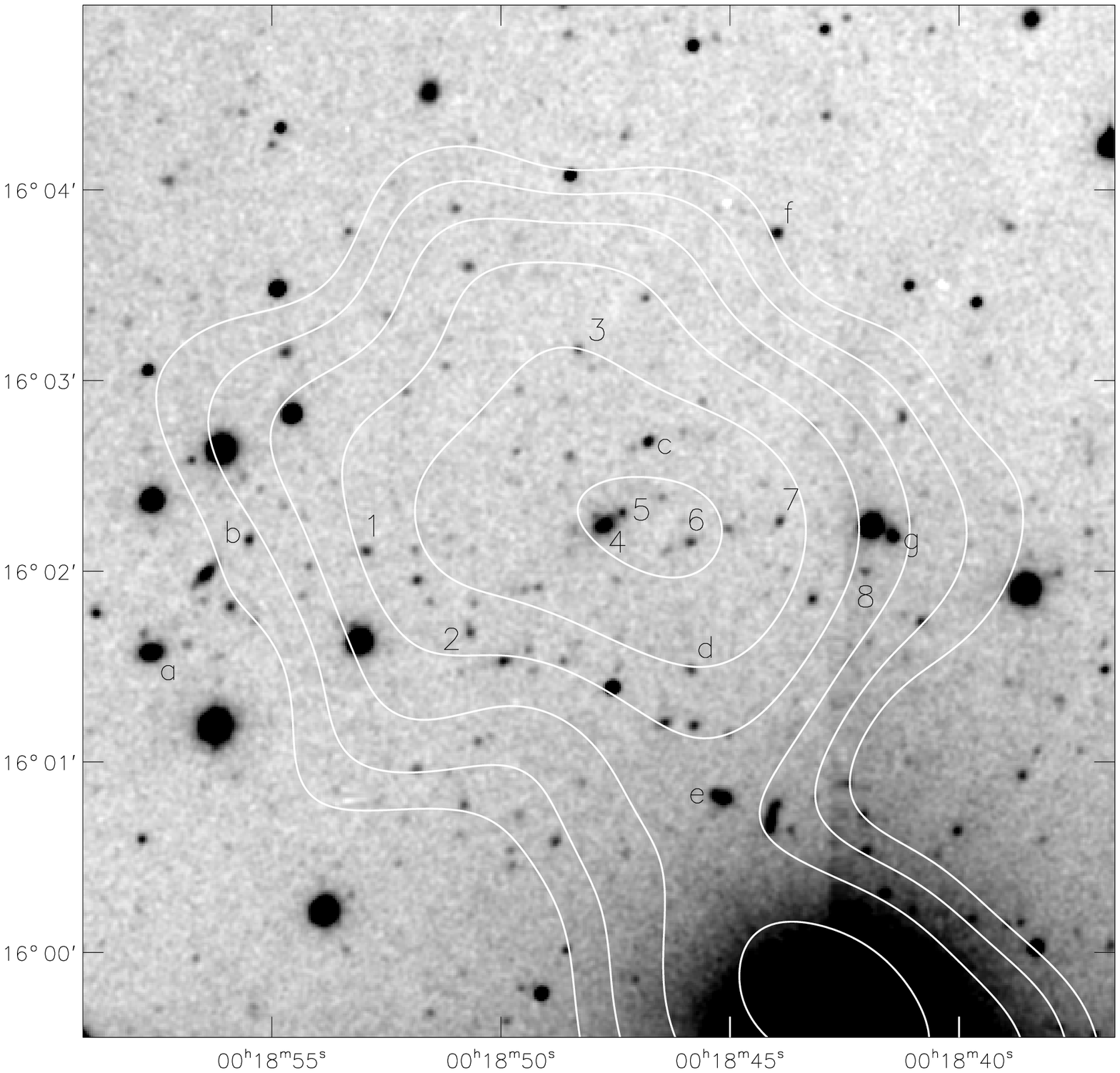}{7in}{0}{100}{100}{-300}{-50}
\end{figure}

\clearpage

\begin{figure}
\plotfiddle{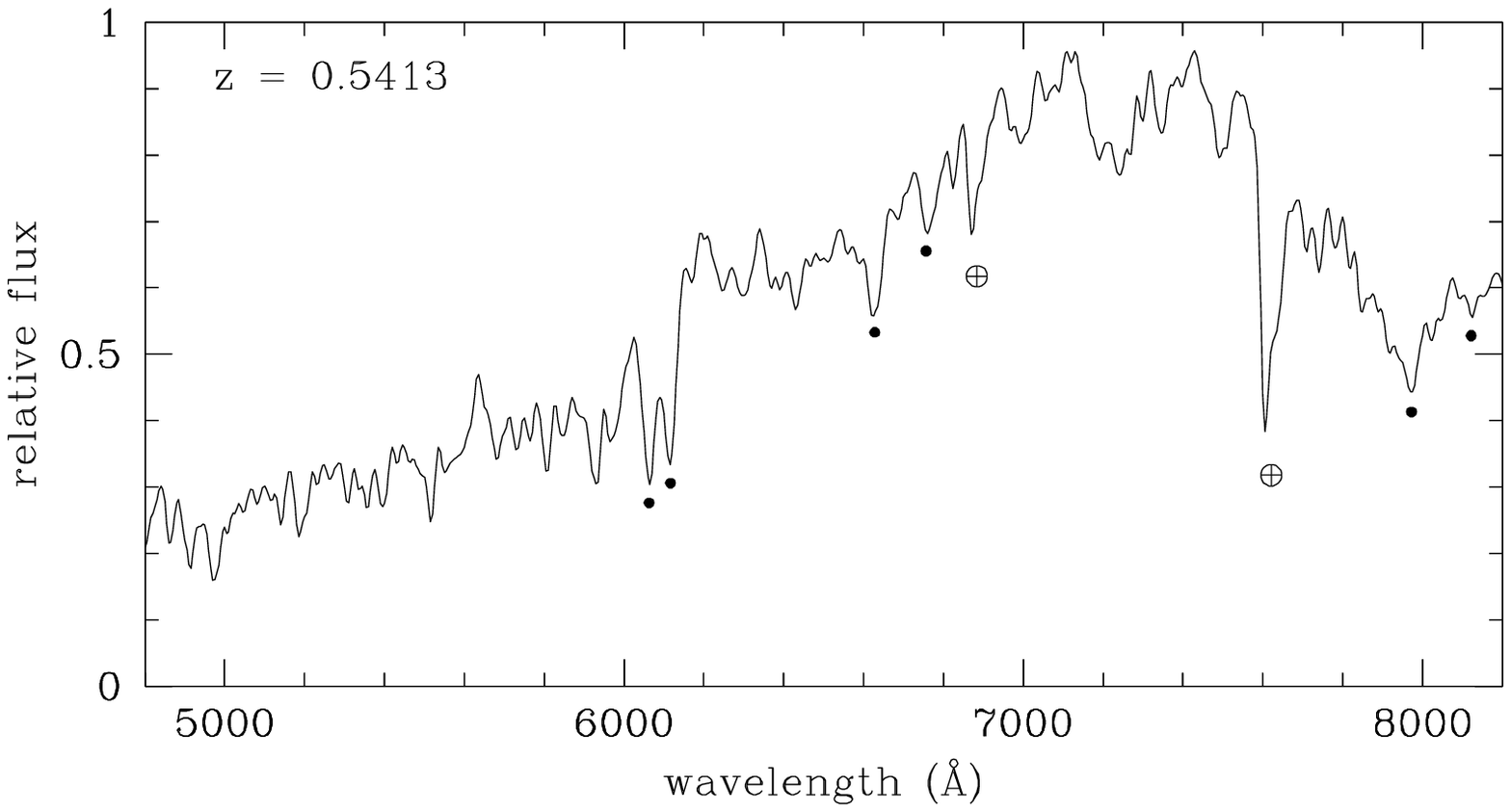}{3in}{0}{100}{100}{-300}{-300}
\end{figure}

\clearpage

\begin{figure}
\plotfiddle{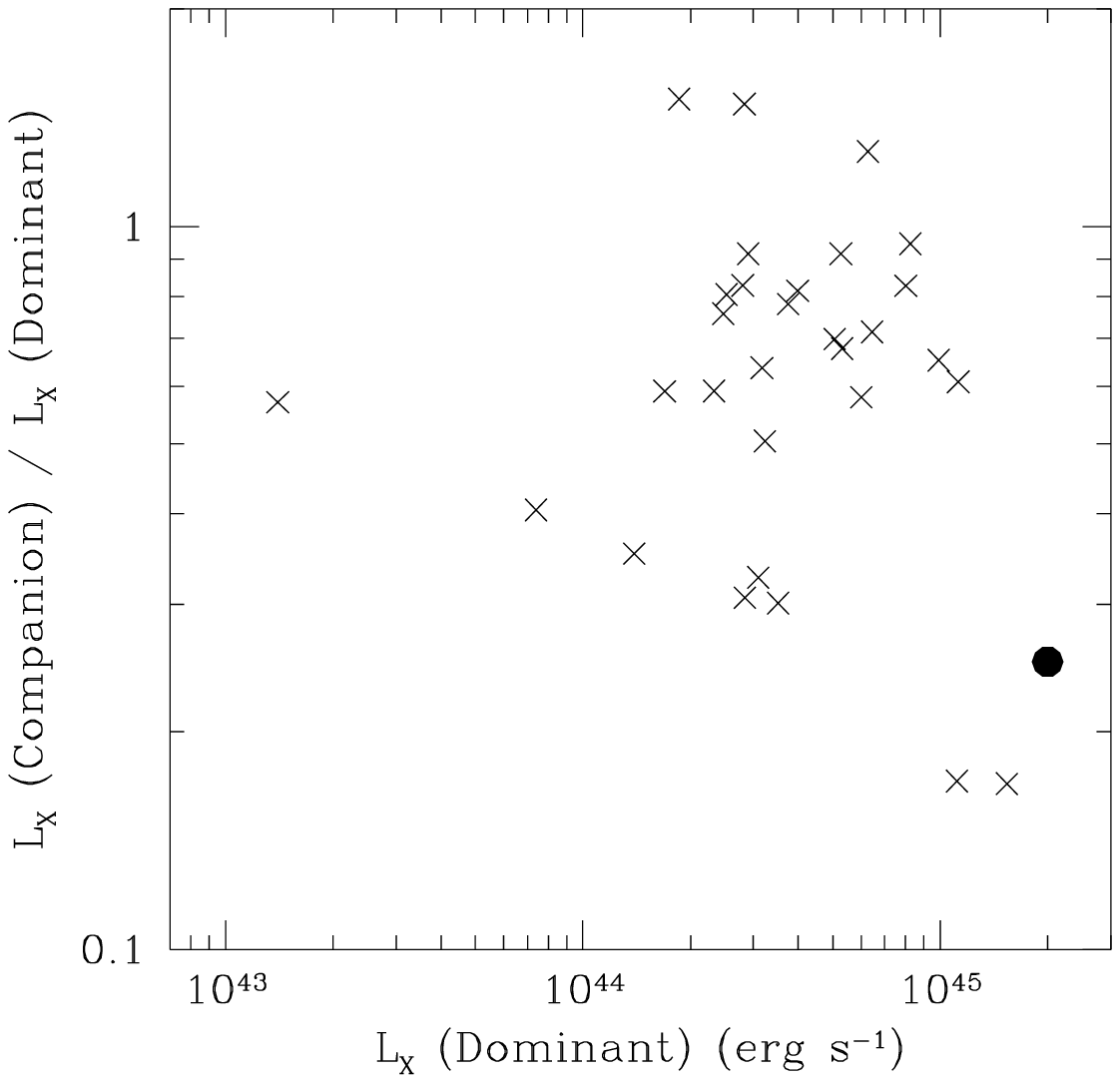}{3in}{0}{100}{100}{-300}{-300}
\end{figure}


\begin{references}

\reference{briel} Briel, U.~G., \& Henry, J.~P.~1993, A\&A, 278, 379

\reference{briel2} Briel, U.~G., \& Henry, J.~P.~1995, A\&A, 309, L9

\reference{car1} Carrera, F.~J., et al.~1991, MNRAS, 249, 698

\reference{car2} Carrera, F.~J., et al.~1993, MNRAS, 260, 376

\reference{cav} Cavaliere,~A., \& {\hbox{Fusco-Femiano}},~R. 1976,
\aap, 49, 137 

\reference{conn1} Connolly, A.~C., Szalay, A.~S., Koo, D., Romer, A.~K.,
Holden, B., Nichol, R.~C., \& Miyaji, T.~1996, \apj, 473, L69

\reference{conn2} Connolly, A.~C., Szalay, A.~S., Romer, A.~K.,
Nichol, R.~C., Holden, B., Koo, D., \& Miyaji, T.~1997, in Proceedings 
of the 18th Texas Symposium in Relativistic Astrophysics (15-20
December 1996), in press (astro-ph/9702025)

\reference{dac} Da Costa, L.~N., et al.~1994, \apj, 424, L1

\reference{ebel} Ebeling, H., Voges, W., B\"ohringer, H., Edge, A.~C.,
Huchra, J.~P., \& Briel, U.~G.~1996, \mnras, 281 799

\reference{edge} Edge, A.~C., Stewart, G.~C., Fabian, A.~C., \&
Arnaud, K.~A.~1990, MNRAS, 245, 559

\reference{dan} Danese, L., De Zotti, G., \& di Tullio, G.~1980, \aap,
82, 322

\reference{fmb} Fabricant, D.~G., McClintock, J.~E., \& Bautz,
M.~W.~1991, \apj, 381, 33

\reference{has} Hasinger, G., Boese, G., Predehl, P., Turner, T.~J.,
Yusaf, R., George, I.~M., \& Rohrbach, G.~1994, Legacy, 4, 40

\reference{jph1} Hughes, J.~P., \& Birkinshaw, M.~1997, \apj, submitted

\reference{jph2} Hughes, J.~P., Birkinshaw, M., \& Huchra, J.~P.~1995,
\apj, 448, L93

\reference{koo} Koo, D.~C.~1981, \apj, 251, L75

\reference{koo} Koo, D.~C.~1986, \apj, 311, 651

\reference{lah} Lahav, O., Edge, A.~C., Fabian, A.~C., \& Putney,
A.~1989, MNRAS, 238, 881

\reference{Mul} Mulchaey, J.~S., Davis, D.~S., Mushotzky, R.~F.,
Burstein, D.~1996, \apj, 456, 80

\reference{neu} Neumann, D.~M., \& B\"ohringer, H.~1997, MNRAS, 289, 123

\reference{nichol} Nichol, R.~C., Briel, U.~G., \& Henry, J.~P.~1994,
MNRAS, 267, 771

\reference{soltan} Soltan, A.~M., Hasinger, G., Egger, R., Snowden,
S., \& Tr\"umper, J.~1996, A\&A, 305, 17

\reference{sto} Stocke, J.~T., Morris, S.~L., Gioia, I.~M., Maccacaro,
T., Schild, R., Wolter, A., Fleming, T.~A., \& Henry, J.~P.~1991,
\apjs, 76, 813

\reference{win} Windhorst, R.~A., Kron, R.~G., \& Koo, D.~C.~1984,
\aaps, 58, 39

\end{references}
\end{document}